# Microfocus small-angle X-ray scattering at SSRF BL16B1[*]

Wen-Qiang Hua(滑文强), Yu-Zhu Wang (王玉柱)[1)], Ping Zhou (周平), Tao Hu (胡涛), Xiu-Hong Li (李秀宏), Feng-Gang Bian (边风刚), Jie Wang (王劼)

*Shanghai Synchrotron Radiation Facility, Shanghai Institute of Applied Physics, Chinese Academy of Sciences, Shanghai, 201204, China*

**Abstract**：Offering high-brilliance X-ray beams on micrometer length scales, the μSAXS at SSRF BL16B1 was established with a KB mirror system for studying small sample volumes, or probing micro-scopic morphologies. The SAXS minimum q value was $0.1nm^{-1}$ with a flux of $1.5 \times 10^{10}$ photons/s. Two position-resolved scanning experimental methods were combined with μSAXS that include STXM and CT. To improve the significant smearing effect in the horizontal direction, an effective and easy-to-use desmearing procedure for two-dimensional SAXS pattern based on the blind deconvolution was developed and the deblurring results demonstrated the good restoration effect for the defocus image. Finally, a bamboo sample was selected for SAXS-CT experiment which illustrated the performance of the μSAXS method.

## 1. Introduction

The interaction of X-rays with in-homogeneities in matter can cause a small deviation from its incident direction, called small-angle X-ray scattering (SAXS) [1]. As a non-invasive technique, SAXS is widely used to probe the micro/nano-scale structure and fluctuations of non-crystalline material which is strongly related to its system properties and functions [2]. The high photon flux and collimation provide by modern synchrotron sources of third generation has made SAXS a unique scattering technique in terms of angular and time resolution, small sample volume, etc. The third generation 3.5 GeV Shanghai Synchrotron Radiation Facility (SSRF) operates in a top-up injection mode at a constant beam current of 240 mA [3]. Among the 7 Phase-I beamlines, BL16B1 is designed to satisfy the needs of a large variety of X-ray scattering experiments [4]. Owing to the excellent photon beam properties of the low-emittance source SSRF, BL16B1 has been optimized for performing in-situ time-resolved small/wide-angle X-ray scattering in both transmission and grazing-incidence geometries (time resolution of the order of sub-seconds). The community of SAXS facility users is rapidly growing in China in recent years [5]. However, our users can't be satisfied with the existing sub-millimeter beam size. An increasing number of researches have been performed with hierarchically structured nanocomposites, and the examples include polymer fibers, biological tissues, samples confined in micro-channels etc [6-8]. Given this, the microfocus beam SAXS (μSAXS) is a suitable and powerful approach to study small scattering volumes. In addition, micro-structural variations in heterogenous samples could be mapped out by scanning the section with high precision. And furthermore, μSAXS method is recently combined with other scanning techniques, such as computed tomography (CT) and scanning transmission X-ray microscopy (STXM) [9, 10]. At present, many beamline in the world has admitted μSAXS as a main technology and a variety of optical focusing systems have been used to focus hard X-rays, such as Kirkpatrick-Baez mirrors (KB mirrors), compound refractive lens, Fresnel zone plate etc. At European Synchrotron Radiation Facility, the ID13 beamline provided the micrometer and sub-micrometer X-ray beams for in-situ time-resolved SAXS/wide-angle X-ray scattering applications [11]. The MiNaxs beamline of PETRA III was dedicated to micro/nano-focused X-ray scattering [12]. As a microfocus beamline in SSRF, BL15U1 was established for hard X-ray micro/nano-spectrochemical analysis and they didn't have a plan to develop the μSAXS method, the limited sample-to-detector distance has become an obstacle [13]. Therefore,

[*] Supported by National Natural Science Foundation of China (Nos. 11505278 and 11675253)

1)E-mail: wangyuzhu@sinap.ac.cn

μSAXS was a key technology urgently needed for the probe of small scattering volumes at BL16B1. The small source size and high flux in SSRF permit us to focus the beam to micrometer using a highly demagnified optics, and obtain sufficient intensity to perform μSAXS. In this case, a new dedicated μSAXS station based on the BL16B1 was proposed and constructed with a KB mirror system. In addition, the microfocus beam grazing incidence SAXS (μGISAXS) was also available at the BL16B1. Furthermore, two position-resolved scanning experimental methods were developed for our users that included SAXS-STXM and SAXS-CT. In this work, we particularly report an effective and easy-to-use desmearing procedure for two-dimensional SAXS pattern to improve the significant smearing effect in the horizontal direction. The SAXS-CT experimental results of a bamboo sample are presented to illustrate the performance of the μSAXS method.

## 2. Beamline optics and μSAXS experimental setup
### 2.1. Beamline optics

The photon source of the BL16B1 was introduced from a bend magnet of SSRF. A scheme of the main optic components of the beamline was shown in Fig. 1 for the μSAXS method. The monochromatic hard X-rays of 5keV to 20keV was delivered by a Si(111) flat double-crystal monochromator (DCM) located at 21m from the source and the energy resolution was $\Delta E/E = 4 \times 10^{-4}$@10keV. A double focusing toroidal mirror was placed 4m downstream from the DCM and was used to focus the X-ray beam to the detector plane. A water-cooled white beam slit was placed at 20m from the source to define the incident beam's receiving angle (1.2mrad*1mrad). Since normal SAXS needed a well collimated and clean beam, three other monochromatic slits were situated at 23m, 35m and 38m to collimate the downstream beam and suppress the upstream stray light.

To perform well the μSAXS experiment, a micro-sized hard X-ray beam is necessary for small volume scattering analysis, while a small beam divergence after the sample will keep the SAXS pattern away from the smearing effect. In theory the signal-to-noise ratio of SAXS experiments stand to benefit from the high photon density of incident hard X-ray beam. Many focusing elements were available at the μSAXS beamline in the world, in consideration of wide energy range of adjustment, low flux loss and adjust convenient, the KB mirrors is a good choice for the μSAXS system at SSRF BL16B1. Thus, when in the μSAXS mode, a KB mirror system was installed at 38.5m from the source (See the top-right corner of Fig. 1). At an energy of 10keV, a focused X-ray spot was achieved at the sample position with a flux of $1.5 \times 10^{10}$photons/s, the beam size was 19.9(±0.1) μm (horizontal) × 13.5(±0.1) μm (vertical) ((See the bottom-left corner of Fig. 1)) and the divergence was about 1.76(±0.05) mrad (horizontal) × 0.55(±0.05) mrad (vertical). The calculated average photon density of focused X-ray was about $5.6 \times 10^7$ photons/s/mm$^2$, which was 474 times greater than that of normal SAXS (Flux: $1.5 \times 10^{11}$photons/s; Beam size: $1600 \times 800 \mu m^2$; Calculated average photon density: $1.2 \times 10^5$photons/s/mm$^2$). To get a clean incident beam and low background data, an anti-scattering pinhole was placed upstream from the sample. As a conventional equipment of the SAXS endstation, a 2m evacuated flight tube (sealed by kapton windows) was placed between the sample and the detector to reduce additional signal due to air scattering. The two-dimensional (2D) SAXS data were recorded using a Mar165 CCD or a Pilatus 200K(DECTRIS) detector. In order to protect the detector from the radiation damage of the direct beam, several beam-stops with different size and shape (circular and rectangle) were available. In addition, there were two beam intensity monitors including a $N_2$ gas ionization chamber before the sample and a photodiode in the beam stop after the sample.

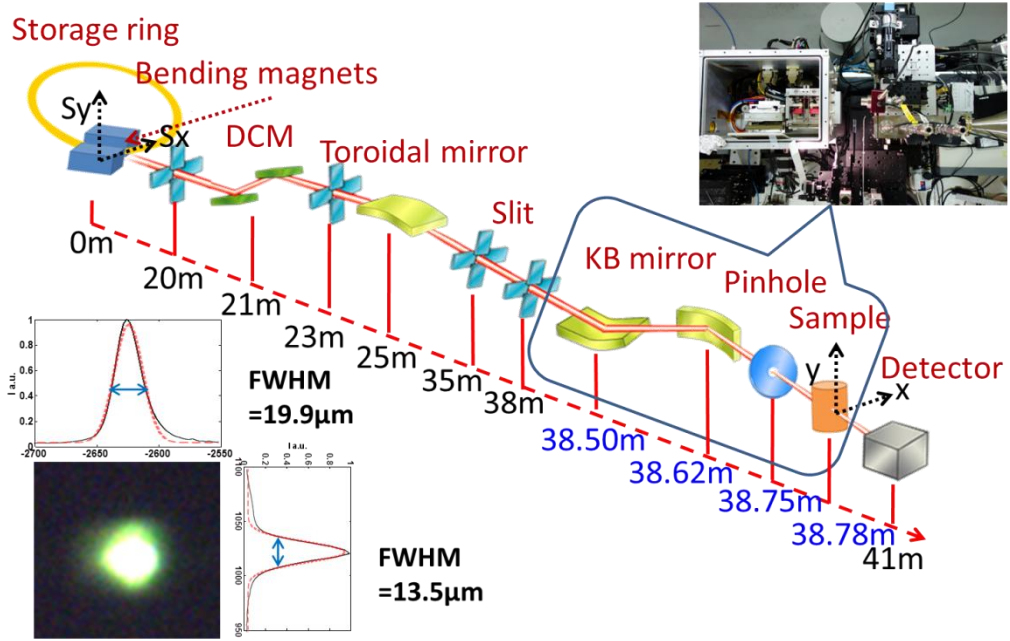

Fig. 1. Schematic map of BL16B1 for the μSAXS method at SSRF.

**2.2. μSAXS experimental modes and data acquisition**

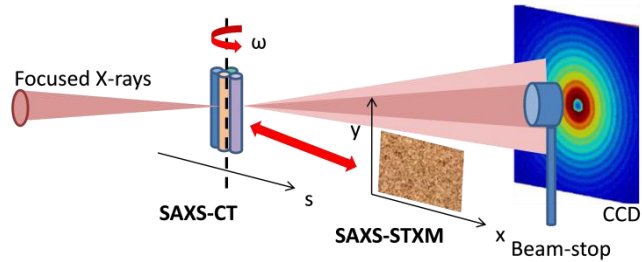

Fig. 2. Experimental set-up for both SAXS-CT and SAXS-STXM measurements.

The SAXS is an averaging technique that access to the nano-scale structures but cannot spatially resolve extended samples. Standard x-ray absorption CT has been used for many years for non-invasive 3D studies, which provides access to volume-resolved information. While providing excellent spatial resolving power, it has an inherent lack of information about the nano-scale structure of the sample [10]. High-resolution scanning techniques like STXM reveal information that typically limited to covering small fields of view in the micrometer range to very thin sample. Full-field two-dimensional x-ray microscopy reveal invaluable information but not down to the nano-scale. [9]. Hence, a combination of SAXS and STXM (or CT) sensitive to nano-scale structures averaged over a spatial area in the square-micron range is suitable for imaging samples of several millimeters to centimeters. The set-up for both SAXS-CT and SAXS-STXM measurements at the BL16B1 are depicted in Fig. 2. For the SAXS-STXM measurements, the incident monochromatic X-ray beam was focused to about $19.9\,\mu m \times 13.5\,\mu m$ at the sample position for the raster scan measurements. To speed up data acquisition, scattering was recorded in a continuous line scan mode with the sample moving along a line of the 2D raster scan while the 2D detector was recording data accordingly. Similarly, for the SAXS-CT measurements, 2D SAXS patterns were collected point by point for each horizontal sample translation $s$, and rotation $\omega$. In addition, the one-dimensional scanning μGISAXS was also developed for our users. All these experimental methods open up a vast field of applications on the investigations of the distribution and orientation of nano-scale structures over extended areas.

**3. Preliminary experiment and data desmearing**

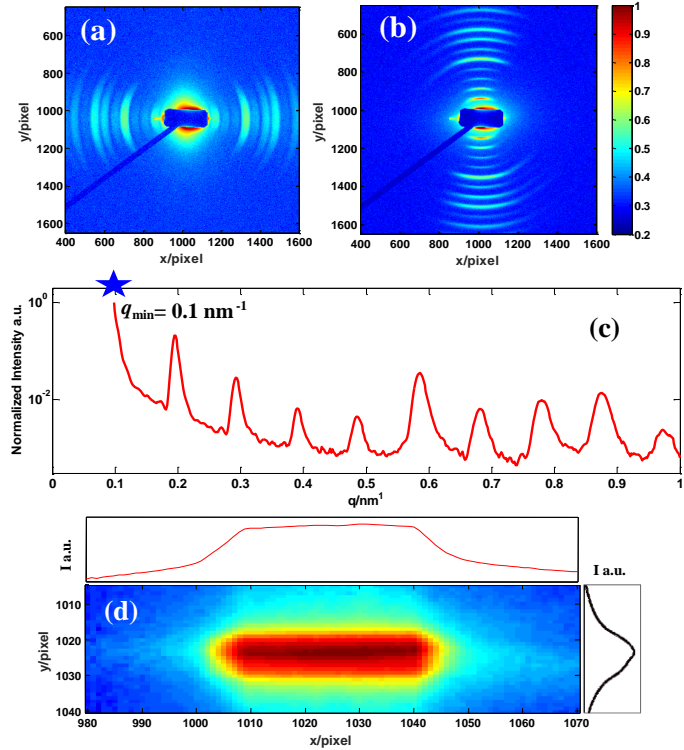

Fig. 3. The μSAXS pattern of a horizontally (a) and vertically (b) placed bull collagen fiber measured at BL16B1 of SSRF; (c) Normalized azimuthal integrated intensity distribution in logarithmic scale marked with $q_{min}$ for (b); (d) The direct beam profile recorded by the detector with aluminum attenuator.

The two-dimensional μSAXS patterns of a widely used standard sample were recorded at BL16B1 of SSRF when the bull collagen fiber was placed horizontally and vertically as shown in Fig. 3(a, b). These intensity profiles were transformed into normalized logarithmic distribution and the good signal-to-noise ratio could be achieved with 100 second exposure time. The SAXS resolution was imposed by the beamstop size and the minimum $q = 2\pi/\lambda \sin 2\theta = 0.1(\pm 0.01)$ nm$^{-1}$ can be obtained in the vertical direction at energy of a 10 keV X-rays as shown in Fig. 3(c). However, the direct beam profile was about $3.5(\pm 0.1) \times 1.1(\pm 0.1)$ mm$^2$ as shown in Fig. 3(d). And because of the smearing effect of the direct beam profile in the horizontal direction, the bandwidth of SAXS streaks was broadened markedly which will lead to the uncertainty of peak width and position.

Instrumental smearing effects, originating from limitations in instrument resolution, can in many instances lead to errors in the SAXS data analysis, or, if data are severely smeared, they can prevent quantitative analysis altogether [14]. With SAXS at modern synchrotron radiation facilities, the smearing effect is usually disregarded because the approximation of an ideal pencil beam is frequently sufficient for the quantitative data analysis. But this may not be appropriate for the tabletop SAXS and μSAXS set-ups [14, 15]. In this case, desmearing of the μSAXS data to correct for instrumental resolution was very necessary.

### 3.1. Analysis of the smearing effect in μSAXS data

The main source of SAXS instrumental smearing originate from finite collimation (the beam profile or divergence), finite detector resolution (pixel size), the wavelength spread, the beam profile at the sample and finite sample thickness [1, 16]. All these aspects contribute to a smearing or blurring effect on the observed scattering pattern, and the definition or sharpness of the SAXS pattern can be reduced which may lead to a misestimate of the regularity of arrangement in material. A number of studies have been performed in order to investigate the smearing effect in SAXS experiment. The true scattering cross section $I_{ideal}(q)$ yields the smeared intensity $I_{obs}(q)$, through a convolution with the resolution function or point-spread functions (PSF) $R(q, q')$,

$$I_{obs} = \int_{q'} R(q,q') I_{ideal}(q') dq' + n(q), \quad (1)$$

where $n(q)$ is the noise introduced in the procedure of image acquisition. An analytical treatment of the resolution functions describes the contributions of different Gaussian functions. For a finite wavelength spread of the radiation, the full width at half-maximum(FWHM) of the Gaussian resolution function is

$$\sigma_\lambda(q') = \frac{1}{2(2\ln 2)^{1/2}} \frac{\Delta\lambda}{\lambda} q'. \quad (2)$$

Besides, the geometric factor $\sigma_{geo}$ also contribute to the total resolution parameter $\sigma_{tot}$,

$$\sigma_{tot}^2(q') = \sigma_\lambda^2(q') + \sigma_{geo}^2, \quad (3)$$

where the geometric contribution include the finite collimation $\sigma_C$ and the detector resolution $\sigma_D$: $\sigma_{geo}^2 = \sigma_C^2 + \sigma_D^2$ [16].

For normal SAXS experiments, the smearing effect due to the sample thickness is really small. A recently published paper demonstrated that the structural parameters calculated from the smeared data sets have little deviation from the ideal ones, which indicates that SAXS data collected in pink-beam mode can be used directly for structural calculations and model reconstructions without a desmearing procedure [17]. Beyond the two factors, the instrument configuration (the profile and divergence of incident beam) and the detector position sensing inaccuracies plays a major role in the smearing of the detected scattering pattern. Most SAXS instruments are designed for optimum resolution in reciprocal space and thus the beam is usually focused onto the detector for minimizing the smearing effect due to beam size and divergence. For micro-beam applications, however, the focus lies preferentially at the sample position, and a compromise has to be found between a minimum beam size and an optimum SAXS resolution. According to Eq. (3), the μSAXS resolution can be estimated by assuming Gaussian functions for the beam profile at the detector as well as for the detector point spread function. Taking into account that the beam size ($3.5\times1.1$ mm$^2$ at the detector position) as shown in Fig. 3(c) is much larger than the detector point spread function (80 μm for the Mar165), the FWHM width of the (Gaussian) resolution function can be approximated by the beam size at the detector plane. Furthermore, we calculated the μSAXS resolution at an energy of 10keV and the scattering vector of $q = 0.6$ nm$^{-1}$ based on Eq. (2, 3), the finite wavelength contribution $\sigma_\lambda = 1.0\times10^{-4}$ nm$^{-1}$, the detector pixel size contribution $\sigma_D = 8.6\times10^{-4}$ nm$^{-1}$, the finite collimation contribution $\sigma_D = 3.7\times10^{-2}$ nm$^{-1}$ (horizontal) and $1.2\times10^{-2}$ nm$^{-1}$ (vertical), so the total μSAXS resolution $\sigma_{tot} = 3.7\times10^{-2}$ nm$^{-1}$ (horizontal) and $1.2\times10^{-2}$ nm$^{-1}$ (vertical) are equal to the finite collimation resolution. Thus, the compound smearing contributions for μSAXS at BL16B1 can be directly evaluated as the image of the direct beam on the detector.

### 3.2. Desmearing procedure for two-dimensional μSAXS data

In order to analyze the SAXS data measured with a significant instrumental smearing, many desmearing methods have been developed since the SAXS was born. They were mainly performed by the iterative desmearing procedure, such as the Lake algorithm (for 1D SAXS pattern) and the Van Cittert method [16]. In general, these methods are somewhat not easy to handle. Besides, using these algorithms for desmearing a large number of data sets are thus not convenient and can be time-consuming. 2D desmearing of centrosymmetric SAXS pattern was developed by the Wiener filtering method, but the limitations for applying this filter was a tedious task [18].

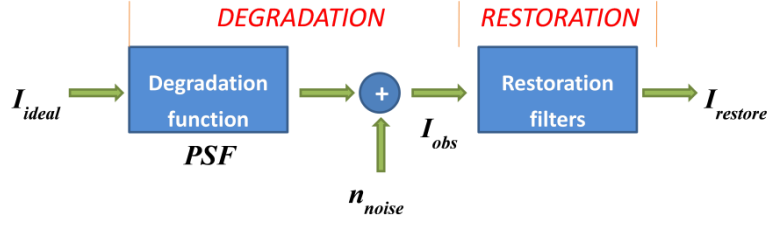

Fig. 4. A model of the image degradation and restoration process.

In computer vision and image processing, the image restoration has being a research focus because of the degradation of image resolution and contrast as shown in Fig. 4. Existing image restoration algorithms, such as Inverse filtering, Wiener filtering and Lucy-Richardson algorithm built on the known PSF. However, in practice PSF was often unknown or uncertain because of the complexity of imaging conditions. As shown in Fig. 3(d), the direct beam profile was recorded by detector with aluminum attenuator, so the PSF for the smeared μSAXS data is still uncertain. If used naively, the direct beam profile can lead to invalid restoration by the classical restorational methods such as Wiener filtering. Therefore, the application of blind deconvolution and restoration on the μSAXS data is very necessary with uncertain PSF in real experiments. Here, we are interested to present a robust and easy-to-use method to desmear 2D SAXS data. The blind deconvolution process could be completed by calling the "deconvblind" function in MATLAB which is similar to the process of the damped Lucy-Richardson algorithm by accelerated convergence [19]. The Lucy–Richardson deconvolution algorithm is one of the most commonly used procedures for image deblurring/enhancement [20]. The algorithm works based on calculating the maximum-likelihood solution for recovering an undistorted image that has been blurred by a known PSF. The undistorted image could be resolved via an iterative process using the following formula

$$I_{resotre(m+1)}(x, y) = I_{restore(m)}(x, y) \left[ I_{obs}(-x,-y) \otimes \frac{PSF(x, y)}{I_{obs}(-x,-y) \otimes I_{restore(m)}(x, y)} \right], \quad (4)$$

where $I_{restore(m)}(x, y)$ is the estimate of the undistorted image in the $m$th iteration and "$\otimes$" indicates the convolution operation. By using the "deconvblind" function in MATLAB, the restoration of μSAXS image $I_{obs}$ (Fig. 3(a, b)) was calculated with the initial PSF (Fig. 3(d)). The restored image and estimated PSF was output in short processing time. And, the deblurring results of the μSAXS defocus image (Fig. 5(b, c)) demonstrated high quality restoration when compared with normal SAXS image (Fig. 5(a)).

In the process of deconvolution for Fig 3(a), a larger digital beamstop was needed for the parasitic scattering near the beamstop which may caused some deviation in the image restoration. Accompanied with the image blind restoration, the ringing effect existed in the restored image and was more obvious near the beamstop and the edge of CCD. However, these deviation didn't influence the analysis of the SAXS $I(q)$ fringe. As shown in Fig. 5(d, e), a fringe comparison was performed and showed good consistency between the restored image $I_{restore}$ of the smeared image $I_{obs}$ and undesmeared normal SAXS image.

Due to the long and narrow beam profile on the CCD, the peak profiles were remarkably broaden in the horizontal direction, but it's virtually invisible in the vertical direction. When the μSAXS data were desmeared, the $I(q)$ fringes were compared with normal SAXS experimental results. The peak positions and widths of $I(q)$ fringes are very important in SAXS data analysis, so the Gaussian function fitting results of three peaks' positions and widths for bull collagen fibers are shown in Table 1, where $\Delta P/P = |P_{obs}-P_{normal}|/P_{normal}$ and $\Delta W/W = |W_{obs}-W_{normal}|/W_{normal}$ are the ratio of peak and width deviation to the normal SAXS value. The peak position and width deviation in the vertical direction was hardly visible for the small resolution function according to Eq. (1), so the desmearing procedure was consequentially unnecessary in the vertical direction. However, the effect of desmearing procedure in the horizontal direction was very clear, the peak width deviations of processed images decreased to regular levels while the peak position deviations were still kept in normal after treatment. Meanwhile, there were some deficiencies. Obviously, the missing SAXS data in low q zone can absolutely obstacle the Guinier analysis in the horizontal direction, which may be circuited by 90 degree rotating of samples. All these results demonstrate the effectiveness and feasibility of this desmearing method. Finally, with this easy-to-use desmearing method, the μSAXS method at BL16B1 has become an entire system.

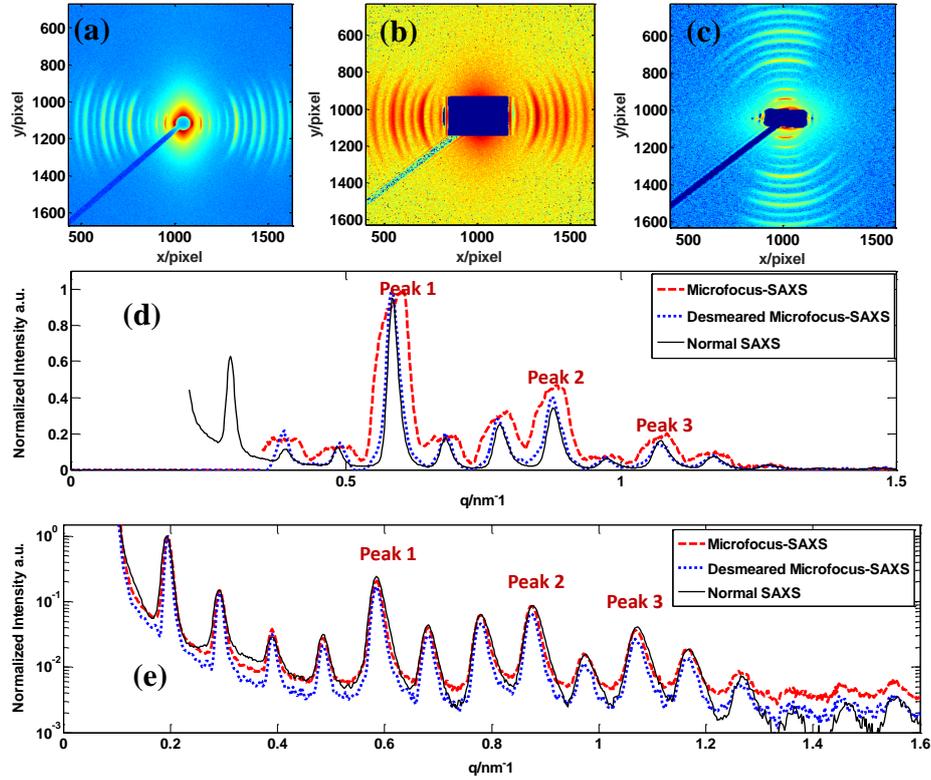

Fig. 5. (a) The normal SAXS image of horizontal placed bull collagen fiber; (b,c) The restored image for Fig. 3(a, b); (d,e) The $I(q)$ fringe comparison of Fig. 3 (a, b), Fig. 5(b, c) and 5(a).

Table 1. The fitting results of peak positions and widths as marked in Fig. 5(d, e) using Gaussian function.

|  |  | Normal SAXS | μSAXS Horizontal Bull |  | μSAXS Vertical Bull |  |
|---|---|---|---|---|---|---|
|  |  | Horizontal Bull | Before desmeared | After desmeared | Before desmeared | After desmeared |
| Peak 1 | $P$ (nm$^{-1}$) | 0.5880 | 0.5871 | 0.5855 | 0.5853 | 0.5853 |
|  | $\Delta P/P$ | - | 0.15% | 0.45% | 0.46% | 0.46% |
|  | $W$ (nm$^{-1}$) | 0.0232 | 0.0837 | 0.0283 | 0.0200 | 0.0176 |
|  | $\Delta W/W$ | - | 2.61 | 0.22 | 0.14 | 0.24 |
| Peak 2 | $P$ (nm$^{-1}$) | 0.8803 | 0.8760 | 0.8766 | 0.8755 | 0.8756 |
|  | $\Delta P/P$ | - | 0.49% | 0.42% | 0.55% | 0.53% |
|  | $W$ (nm$^{-1}$) | 0.0290 | 0.0737 | 0.0345 | 0.0243 | 0.0223 |
|  | $\Delta W/W$ | - | 1.54 | 0.19 | 0.16 | 0.23 |
| Peak 3 | $P$ (nm$^{-1}$) | 1.0754 | 1.0725 | 1.0709 | 1.0704 | 1.0706 |
|  | $\Delta P/P$ | - | 0.18% | 0.42% | 0.46% | 0.50% |
|  | $W$ (nm$^{-1}$) | 0.0325 | 0.0701 | 0.0368 | 0.0283 | 0.0270 |
|  | $\Delta W/W$ | - | 1.16 | 0.13 | 0.13 | 0.17 |

## 4. SAXS-CT for Bamboo

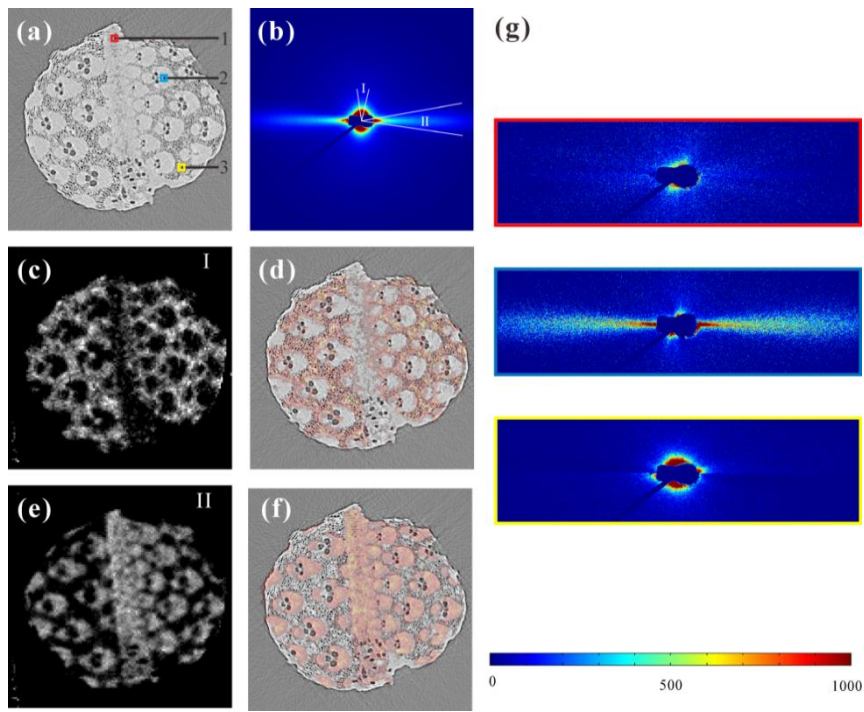

Fig. 6. SAXS-CT for the bamboo sample. (a) The X-ray absorption micro-CT slice of the bamboo sample (1: Adhesive; 2: Vascular bundle; 3: Parenchyma cell); (b) Two-dimensional SAXS pattern of the bamboo sample. (c) The scattering tomographic slice reconstructed by integrating the region I of (b). (d) The overlap between (c) and (a). (e) The scattering tomographic slice reconstructed by integrating the region II of (b). (f) The overlap between (e) and (a). (g) The reconstructed 2D SAXS patterns for three locations in the bamboo sample, indicated by the red, blue and yellow squares.

To illustrate the performance of the μSAXS system at SSRF BL16B1, a bamboo sample was selected for SAXS-CT experiment as shown in Fig.6. The experiment was performed with 117 translations at a step size of 30 μm and 30 projection angles over $180^o$, 20s exposure time per frame. A total of 3510 SAXS patterns were collected in total 24 hours. In fig. 6(c) and (e), by selecting different integral region of the scattering signals, the different tomographic slice of the sample could be reconstructed. As shown in Fig. 6(a), a normal CT reconstruction for this bamboo sample was also performed at SSRF BL13W1 which could further verified the SAXS-CT results by overlap Fig. 6(a) and (c) or (e). Furthermore, after reconstructing all the tomographic slices for each $q$, the 2D SAXS pattern at each location in the sample can be mapped. Fig. 6(g) shows the differences of the reconstructed 2D SAXS patterns for three different regions in the bamboo sample.

## 5. Conclusion

To probe small scattering volumes, a μSAXS instrument has been established with a KB mirror system at SSRF BL16B1 and the high-brilliance X-ray beams could be achieved on micrometer length scales. Two position-resolved scanning experimental methods (STXM and CT) were also developed for our users which could be combined with μSAXS and μGISAXS methods. Besides, we particularly report an effective and easy-to-use desmearing procedure for two-dimensional SAXS pattern to improve the significant smearing effect in the horizontal direction. The application of this blind deconvolution procedure on the smeared SAXS pattern of the bull collagen fiber demonstrated the good restoration effect for the defocus blurred image in short computation time. Furthermore, this desmeared procedure is also expected to be useful for data processing of tabletop SAXS and μSAXS. Finally, the SAXS-CT experimental results of a bamboo sample were presented to illustrate the performance of the μSAXS method. The long-term goal is to develop a user-friendly operation interface and proceed the online data analysis.

*The authors would like to thank Dr. Xu-Ke Liu for helpful discussions.***References**

1. B. R. Pauw, J. Phys.: Condens. Matter, **26** (38): 239501 (2014)
2. T. Narayanan, *Synchrotron Small-Angle X-Ray Scattering*, (Berlin: Springer, 2008), p. 900
3. H. J. Xu, Z. T. Zhao, Nucl. Sci. Tech., **19** (1): 1-6 (2008)
4. F. Tian, X. H. Li, Y. Z. Wang et al, Nucl. Sci. Tech., **26**: 030101 (2015)
5. Z. H. Li, Z. H. Wu, G. Mo et al, Instrum. Sci. Technol., **42** (2):128-141 (2014)
6. O. Paris, C. H. Li, S. Siegel et al, J. Appl. Crystallogr., **40**: s466–s470 (2006)
7. G. Q. Zheng, Z. Jia, X. Liu et al, Polym. Eng. Sci., **52** (4): 725-732 (2012)
8. L. Wang, W. Yang, S. He et al, Mater. Today Commun., **4** (11): 22-34 (2015)
9. O. Bunk, M. Bech, T. H. Jensen et al, New J. Phys., , **11** (12): 123016 (2009)
10. T. H. Jensen, B. Martin, B. Oliver et al, Phys.Med. Biol., **56** (6): 1717-1726 (2011)
11. C. Riekel, M. Burghammer, R. Davies, IOP Conf. Ser. Mater. Sci. Eng., **14**: 012013 (2006)
12. A. Buffet, A. Rothkirch, R. Döhrmann et al, J. Synchrotron Rad., **19**: 647-653 (2012)
13. L. L. Zhang, S. Yan, S. Jiang et al, Nucl. Sci. Tech., **26**: 060101 (2015)
14. J. Bergenholtz, J. Ulama, M. Z. Oskolkova et al, J. Appl. Crystallogr., **49**: 47-54 (2016)
15. N. Stribeck, U. Nöchel. J. Appl. Crystallogr., **41**: 715-722 (2008)
16. T. Vad, W. F. C. Sager, J. Appl. Crystallogr., **44**: 32-42 (2011)
17. W. J. Wang, E. V. Shtykova, V. V. Volkov et al, J. Appl. Crystallogr., **48**: 1935-1942 (2015)
18. V. L. Flanchec, D. Gazeau, J. Taboury et al, J. Appl. Crystallogr., **29**: 110-117 (1996)
19. R. C. Gonzalez, R. E. Woods, B. R. Masters, J. Biomed. Opt., **14**: 331-333 (2009)
20. S. A. Hojjatoleslami, M. R. N. Avanaki, A. G. Podoleanu, Appl. Opt., **52** (23): 5663-5670 (2013)